\documentclass[twocolumn]{aastex62}

\received{\today}
\revised{...}
\accepted{...}
\submitjournal{ApJL}

\shorttitle{Quasi-periodic pulsations in the most powerful {solar} flare of Cycle 24}
\shortauthors{Kolotkov et al.}

\begin{document}

\title{Quasi-periodic pulsations in the most powerful {solar} flare of Cycle 24}

\correspondingauthor{Dmitrii Y. Kolotkov}
\email{D.Kolotkov@warwick.ac.uk, D.Kolotkov.1@warwick.ac.uk}

\author{Dmitrii Y. Kolotkov}
\affiliation{Centre for Fusion, Space and Astrophysics, Department of Physics, University of Warwick, CV4 7AL, UK}

\author{Chloe E. Pugh}
\affiliation{Centre for Fusion, Space and Astrophysics, Department of Physics, University of Warwick, CV4 7AL, UK}

\author{Anne-Marie Broomhall}
\affiliation{Centre for Fusion, Space and Astrophysics, Department of Physics, University of Warwick, CV4 7AL, UK}
\affiliation{ Institute of Advanced Study, University of Warwick, 
	Coventry, CV4 7HS, UK}

\author{Valery M. Nakariakov}
\affiliation{Centre for Fusion, Space and Astrophysics, Department of Physics, University of Warwick, CV4 7AL, UK}
\affiliation{St. Petersburg Branch, Special Astrophysical Observatory, Russian Academy of Sciences, 196140, St. Petersburg, Russia}

\begin{abstract}

{Quasi-periodic pulsations (QPP) are common in solar flares and are now regularly observed in stellar flares.} We present the detection of two different types of QPP signals in the thermal emission light curves of the X9.3 class solar flare SOL2017-09-06T12:02, which is the most powerful flare of Cycle 24. The period of the shorter-period QPP drifts from about 12 to 25 seconds during the flare. The observed properties of this QPP are consistent with a sausage oscillation of a plasma loop in the flaring active region. The period of the longer-period QPP is about 4 to 5 minutes. {Its properties are compatible with standing slow magnetoacoustic oscillations, which are often detected in coronal loops. For both QPP signals, other mechanisms such as repetitive reconnection cannot be ruled out, however.}
{The studied solar flare has an energy in the realm of observed stellar flares, and the fact that there is evidence of a short-period QPP signal typical of solar flares along with a long-period QPP signal more typical of stellar flares suggests that the different ranges of QPP periods typically observed in solar and stellar flares is likely due to observational constraints, and that similar physical processes may be occurring in solar and stellar flares.}
\end{abstract}


\section{Introduction} \label{sec:intro}

The X9.3 class solar flare that occurred on 2017 September 6 is the most powerful since 2005, and hence is the largest flare observed with the latest generation of instruments and in Cycle 24. Using the relationship between the GOES soft X-ray flux in the 1--8\,\AA\ waveband and the flare energy calculated from the total solar irradiance found by \citet{2011A&A...530A..84K}, X9.3 class corresponds to an energy of around $10^{32}$\,erg. The energy of this flare is therefore in the realm of typically observed stellar flare energies, and solar flares such as this are useful for bridging the energy gap between solar and stellar flares \citep{2015EP&S...67...59M}.

Quasi-periodic pulsations (QPP) are a common feature of solar flares, detected in all observational wavebands, from radio to gamma-rays, in all phases of the flare \citep[see][for recent comprehensive statistical studies]{2016ApJ...833..284I,2017A&A...608A.101P}, and in both thermal and nonthermal emissions \citep[see][respectively]{2013ApJ...777..152S,2010SoPh..267..329K}.
{QPP are a transient phenomenon, hence they tend to be seen in part of the flare light curve, rather than being seen throughout the whole flare.}
The specific values of QPP periods range from a fraction of a second to several tens of minutes \citep[e.g.][]{2016SSRv..200...75N, 2016SoPh..291.3143V}, and hence are likely to be associated with several different physical mechanisms \citep[see][for a recent review]{2018SSRv..214...45M}. It has been established that QPP could be caused by several groups of mechanisms, including {MHD-wave-driven} and spontaneous magnetic reconnection. Revealing these mechanisms {is still an active research area.}

QPP are also detected in stellar flares in the same observational channels as solar flares (e.g. \citeauthor{2005A&A...436.1041M} \citeyear{2005A&A...436.1041M};  \citeauthor{2009ApJ...697L.153P} \citeyear{2009ApJ...697L.153P}; \citeauthor{2013ApJ...778L..28S} \citeyear{2013ApJ...778L..28S}; \citeauthor{2016ApJ...830..110C} \citeyear{2016ApJ...830..110C}, for X-rays, and e.g. \citeauthor{1989A&A...220L...5G} \citeyear{1989A&A...220L...5G}; \citeauthor{2004AstL...30..319Z} \citeyear{2004AstL...30..319Z}, for radio). Stellar QPP are usually detected in flares on active red dwarfs, but are also observed on Sun-like stars. In addition, QPP are detected in stellar flares in the white light (WL) emission \citep[e.g.][]{2003A&A...403.1101M, 2013ApJ...773..156A}. A recent statistical study demonstrated that a significant fraction of WL flares detected with Kepler have QPP \citep{2016MNRAS.459.3659P}.

The WL stellar flares have recently attracted major attention in the context of devastating superflares that can strongly affect the habitability of the planets orbiting a flaring star. Naturally there appears a question of whether the Sun is capable of producing a superflare, and if so what the superflare occurrence rate is \citep[e.g.][]{2012Natur.485..478M, 2017ApJ...851...91N, 2018ApJ...853...41T}. In solar flares the enhancement of the WL emission is usually very weak, with values typically less than 0.01\% of the total irradiance. In contrast, in strong stellar WL flares the increase in the star's irradiance is comparable to its irradiance in the quiet period. Moreover, some stellar WL flares do not have a significant X-ray flux \citep[e.g.][]{1991ARA&A..29..275H}. Thus it is not clear whether stellar WL superflares and solar flares are produced by the same physical mechanisms, and whether the results of stellar flare studies could be scaled down to the Sun. The detection of QPP in stellar WL flares with properties similar to those of QPP in solar flares, in particular with damping patterns \citep{2016MNRAS.459.3659P, 2016ApJ...830..110C}, and multiple periods \citep{2015ApJ...813L...5P, 2018MNRAS.tmp...77D}, suggests that the mechanisms could be the same.

So far, the most powerful solar flare with a QPP pattern detected is an X14.4 flare \citep{2006A&A...460..865M}. The flare had QPP with periods ranging from one to five minutes, detected in the radio and hard X-ray emission. In this letter we present the second most powerful solar flare with detected QPP, which is the most powerful flare of Cycle 24. {The letter is organised as follows:} observations used for the analysis are described in Sec.~\ref{sec:obs}, Sec.~\ref{sec:methods} gives a methodology used for processing the observational time series, the obtained results are summarised in Sec.~\ref{sec:res} and discussed in Sec.~\ref{sec:disc}.

\section{Observations}
\label{sec:obs}

\begin{figure*}
	\centering
	\includegraphics[width=0.8\linewidth]{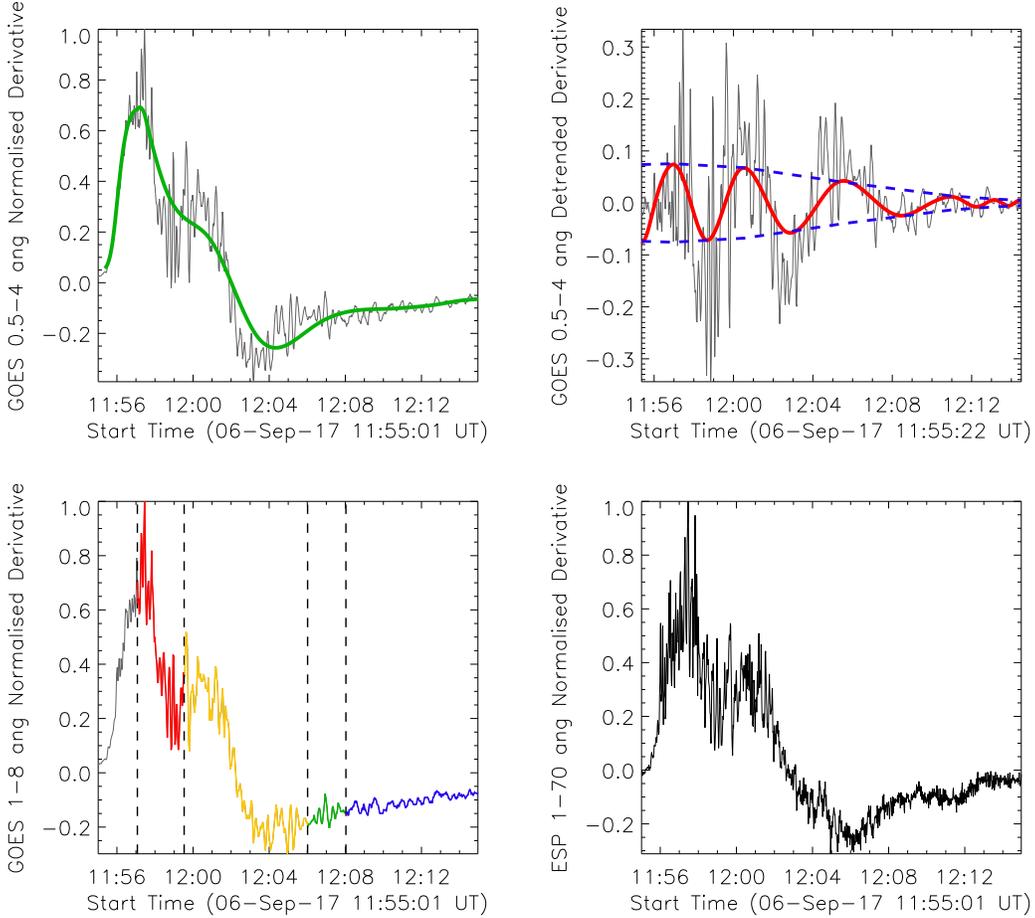}
	\caption{Time derivatives of the fluxes measured by GOES/XRS ({both left panels}) and SDO/EVE/ESP (bottom right) during the impulsive phase of the flare, normalised to their maximum value. The green line in the top left panel shows an overall trend of the 0.5--4\,\AA\ signal, detected with {the empirical mode decomposition (EMD) method}. The top right panel shows the signal with the overall trend subtracted, and an intrinsic mode (the red line) found to be statistically significant in the detrended signal by EMD (see Fig.~\ref{fig:long spectra}).
		The blue dashed lines show a Gaussian envelope of the intrinsic mode (see Sec.~\ref{sec:res} for details).	
		Different colours in the bottom left panel illustrate the sections of the light curve, used for further analyses (see Figs.~\ref{fig:short_spectra_1} and \ref{fig:short_spectra_2} and Table~\ref{table}).}
	\label{fig:flare}
\end{figure*}

{The flare analysed in this letter occurred on 2017 September 6 and peaked at 12:02:00\,UT. It originated from the active region NOAA\,12673} and is the most powerful flare of Cycle 24. 

In this study we used data from two instruments that observed the flare (see Fig.~\ref{fig:flare}), which are the \emph{Geostationary Operational Environmental Satellite's} X-ray sensor (GOES/XRS), and the \emph{Extreme ultraviolet SpectroPhotometer} (ESP) channel of the \emph{Extreme ultraviolet Variability Experiment} (EVE) aboard the \emph{Solar Dynamics Observatory} (SDO). GOES/XRS makes Sun-as-a-star observations of soft X-ray flux in the 1--8\,\AA\ and 0.5--4\,\AA\ wavebands with a cadence of 2.048\,s, while EVE/ESP covers 1--70\,\AA\ wavelengths with a cadence of 0.25\,s. The uncertainties on the XRS and ESP data were estimated using the same approach as \citet{2017A&A...608A.101P}. 

\section{Data analysis}
\label{sec:methods}
\subsection{Periodogram-based analysis}
\label{sec:fourier}
Stationary QPP, i.e. those with constant periods, were studied with the analysis technique described in detail in \citet{2017A&A...602A..47P}, which is based upon the work of \citet{2005A&A...431..391V} and is outlined below. The technique assesses the statistical significance of peaks in a periodogram, accounting for both data uncertainties and coloured noise, which results in the Fourier power spectral density $S$ and the frequency $f$ being connected as
\begin{equation}
\label{eq:col_noise_fourier}
S\propto f^{-\alpha},
\end{equation}
where the power law index $\alpha$ determines the \lq\lq colour\rq\rq\ of noise. 


\begin{figure*}
	\centering
	\includegraphics[width=0.8\linewidth]{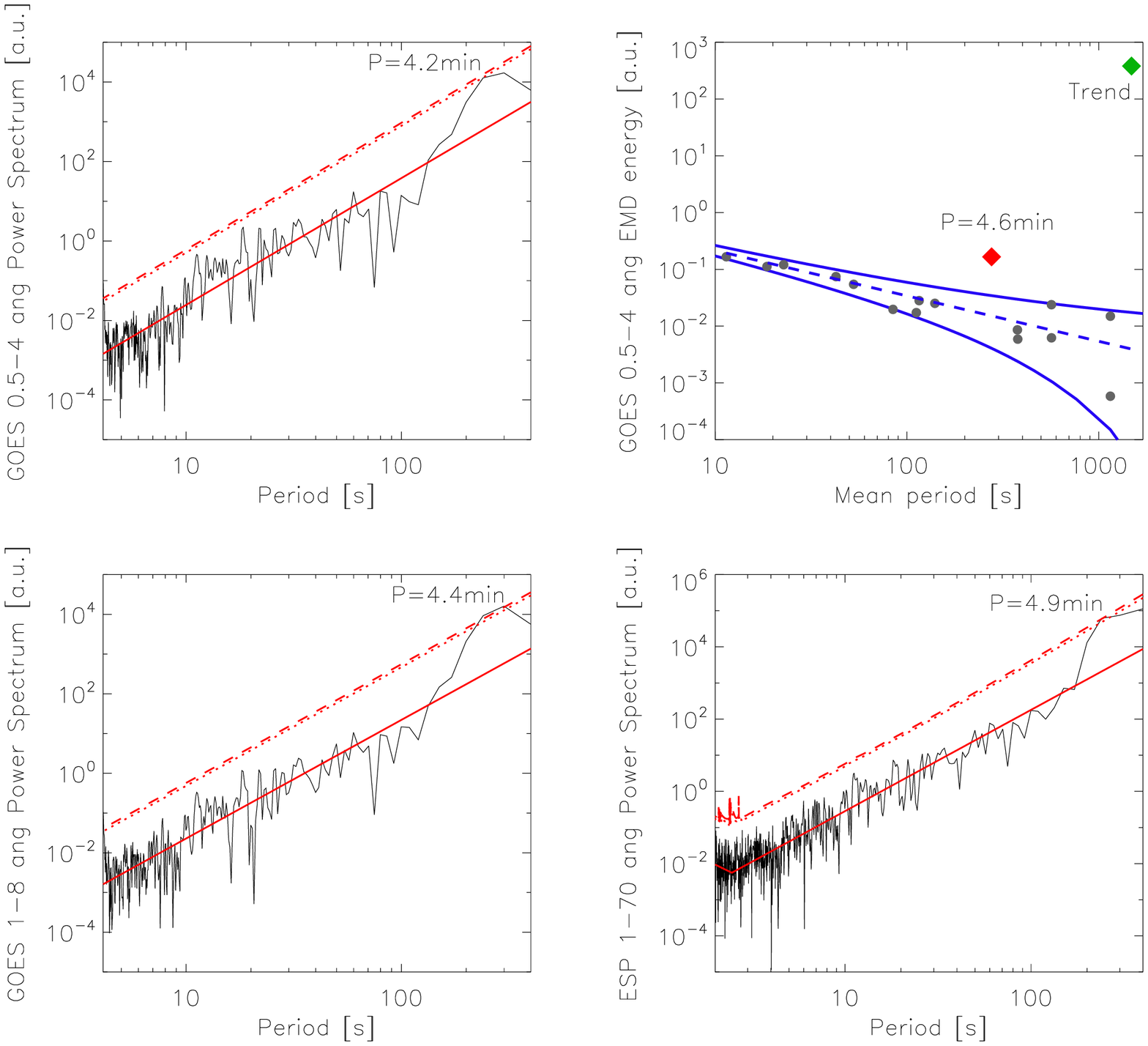}
	\caption{The periodograms of the GOES/XRS (top {left} and bottom left) and SDO/EVE/ESP (bottom right) signals shown in Fig.~\ref{fig:flare}, where the red solid lines are broken power law fits to the spectra, and the red dotted and dashed lines represent the 95\% and 99\% confidence levels, respectively. Top right: EMD spectrum, i.e. the dependence of the total energy of an intrinsic mode upon the mean period, in the detrended 0.5--4\,\AA\ signal (see Fig.~\ref{fig:flare}). The blue solid lines show the limits of the 99\% confidence interval, inside which the modes belong to noise. The mean energy is determined by Eq.~(\ref{eq:col_noise_emd}) and is shown by the blue dashed line. The intrinsic modes associated with noise are shown by the black circles, the mode with statistically significant properties and flare trend are shown by the red and green diamonds, respectively.}
	\label{fig:long spectra}
\end{figure*}

\begin{figure*}
	\centering
	\includegraphics[width=0.4\linewidth]{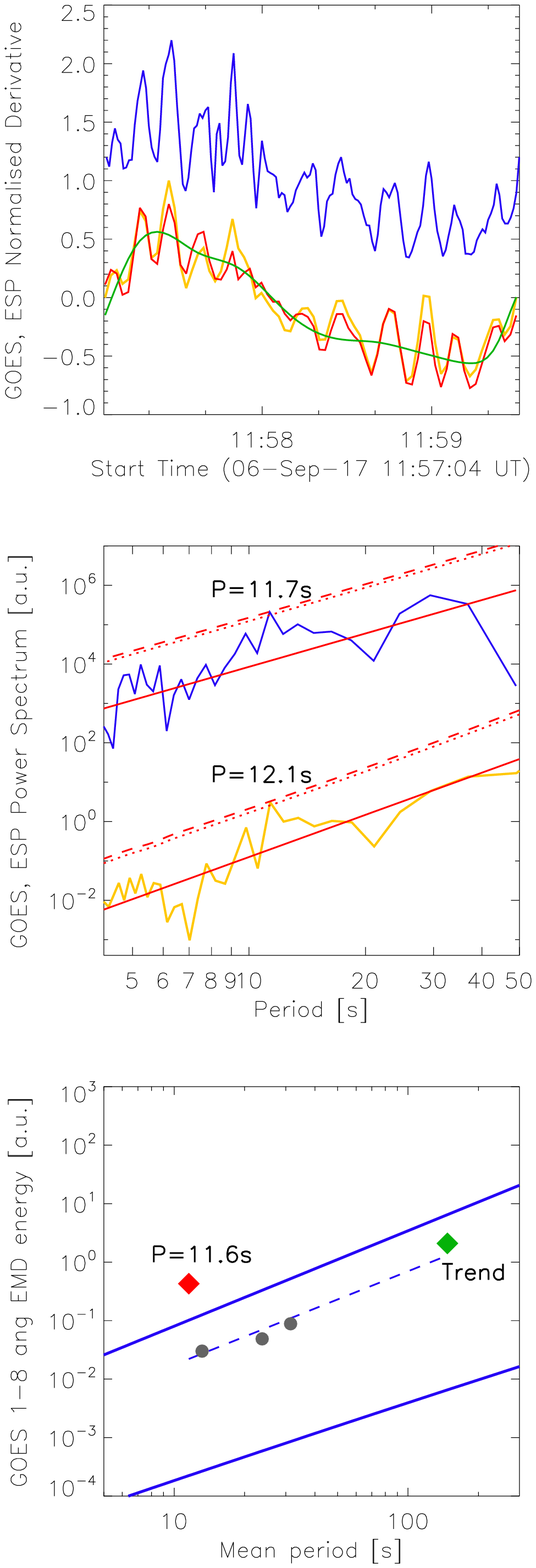}
	\includegraphics[width=0.4\linewidth]{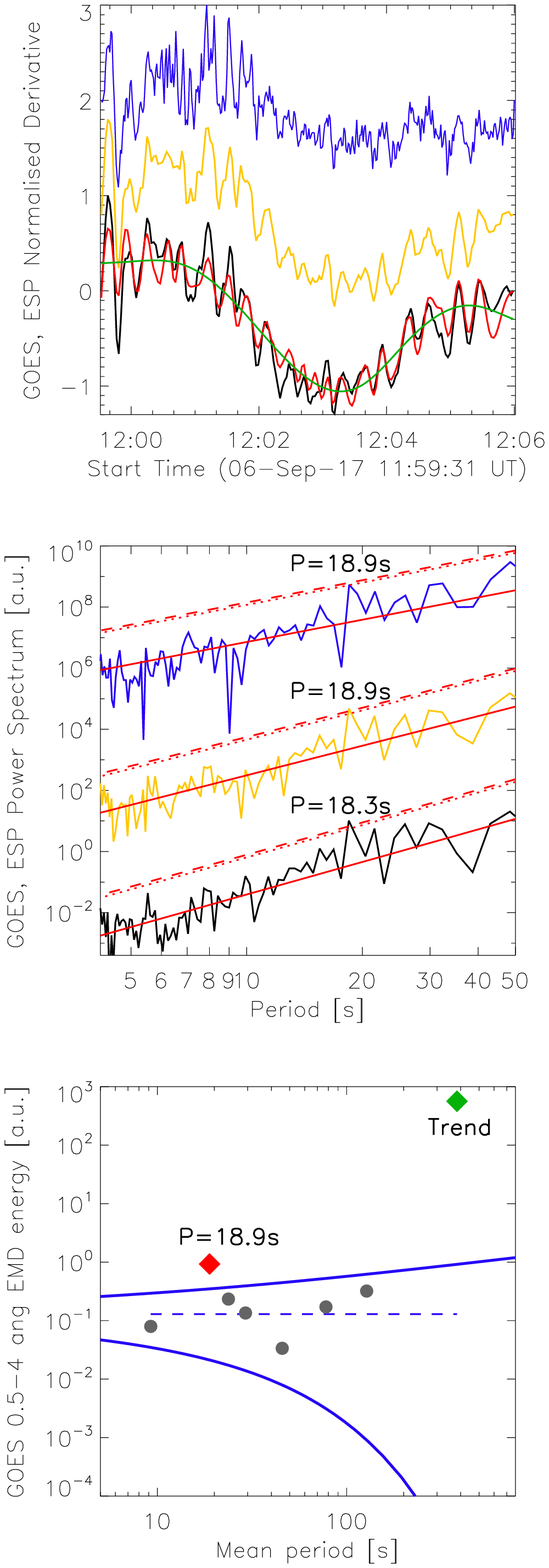}
	\caption{Similar to Figs.~\ref{fig:flare} and \ref{fig:long spectra}, but for the flare sections, starting from 11:57:04\,UT (left) and 11:59:31\,UT (right), {and with a linear interpolation between the start and end values subtracted (see Sec~\ref{sec:fourier} for details)}. Only those light curves with a significant QPP signal are included in these plots. Signals at 1--8\,\AA, 0.5--4\,\AA, and 1--70\,\AA\ wavebands and their power spectra are shown in the top and middle panels by the {yellow}, black, and blue lines, respectively. The green and red lines in the top panels show the overall trends and significant oscillations detected in the corresponding signals with EMD, respectively. All the curves in the top and middle panels were normalised to their maximum values and shifted upwards or downwards for a better visualisation. Bottom panels show the EMD spectra of 1--8\,\AA\ (left) and 0.5--4\,\AA\ (right) signals. The notations are similar to those in Fig.~\ref{fig:long spectra}.}
	\label{fig:short_spectra_1}
\end{figure*}

\begin{figure*}
	\centering
	\includegraphics[width=0.4\linewidth]{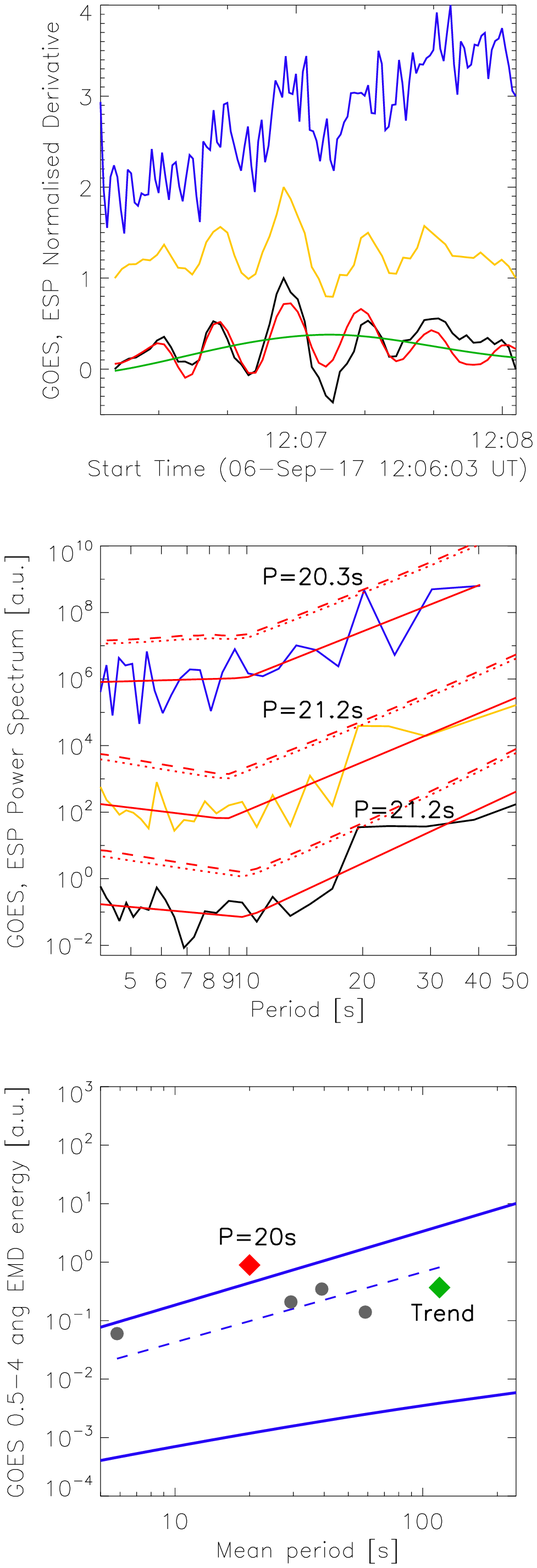}
	\includegraphics[width=0.4\linewidth]{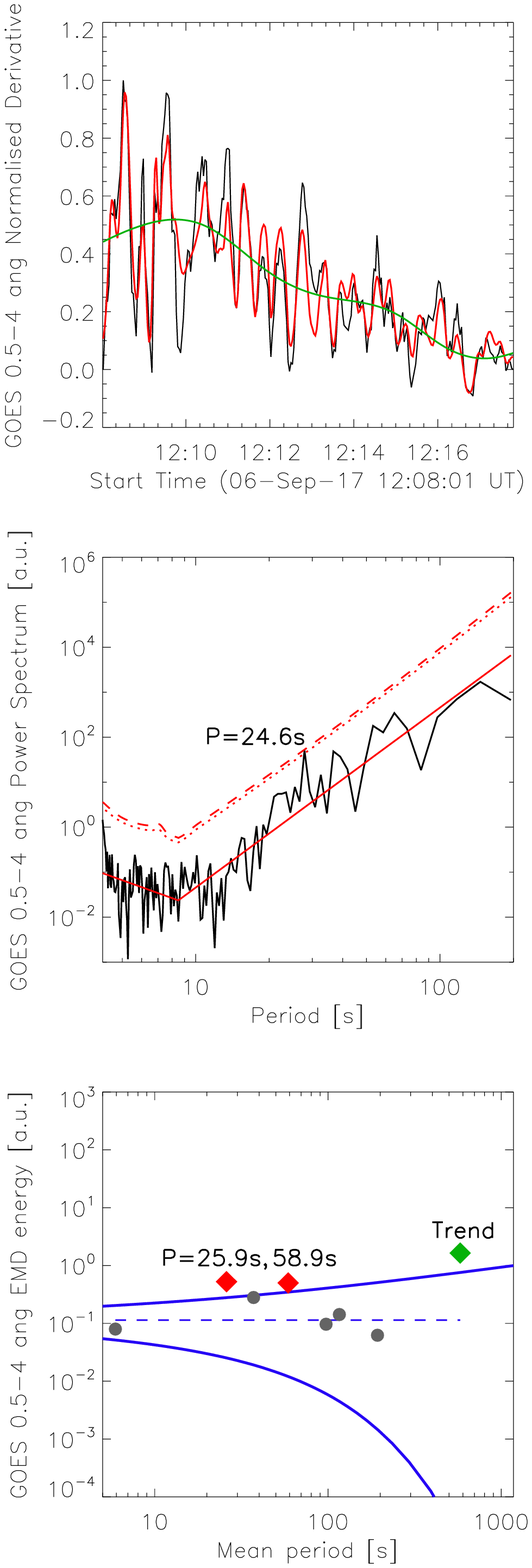}
	\caption{Similar to Fig.~\ref{fig:short_spectra_1}, but for the flare sections, starting from 12:06:03\,UT (left) and 12:08:01\,UT (right). The red line in the top right panel shows a superposition of the two modes found to be significant in the EMD spectrum.}
	\label{fig:short_spectra_2}
\end{figure*}

In addition to the analysis of the entire flare light curves as shown in Figs.~\ref{fig:flare} and \ref{fig:long spectra}, the light curves were manually trimmed to focus on shorter time intervals that were treated separately as potentially containing QPP signals (see Figs.~\ref{fig:flare}, \ref{fig:short_spectra_1}, and ~\ref{fig:short_spectra_2}). Periodograms were then calculated separately for each of the light-curve sections, and a broken power law model was fitted to the resulting periodograms. The model uncertainties at each frequency bin were then factored into the calculation of the 95\% and 99\% significance levels in the manner described by \citet{2017A&A...602A..47P,2017A&A...608A.101P}.


The start and end times of the signal were altered by cutting off or adding individual data points to maximise the height of any peak in the periodogram that could correspond to a potential QPP signal. The combination producing the maximum signal-to-noise ratio was kept. This process is necessary since the signals can be short-lived, meaning that even minor changes to the length of the data can affect the signal-to-noise ratio. Changing the length of the data included in the analysis also changes the resolution of the frequency bins, preventing the case where the signal, by chance, lies between two sampled frequencies.

{For the final step prior to calculating the periodogram, the start and end values of these multiple flare sections were equalised to each other by subtracting a linear interpolation between them. Since calculating the periodogram assumes that the time series data is cyclic, doing this removes the apparent discontinuity between the start and end times that can introduce false signal into the power spectrum. The subtraction of a linear trend in the time domain does not alter the probability distribution of values in the power spectrum, hence it is compatible with the significance testing method used in this letter.}

\subsubsection{Time derivative data}
As shown by previous studies, QPP can be more easily detected in the time derivative of soft X-ray (SXR) observations than in the raw observations themselves \citep[e.g.][]{2015SoPh..290.3625S, 2016ApJ...827L..30H, 2017ApJ...836...84D}. This is because the change in SXR flux during the impulsive phase of a flare, where QPP are predominantly observed, is often vastly greater in amplitude than the QPP being searched for. Since taking the derivative can be thought of as a form of detrending, its impact is taken into account too.
\citet{2017A&A...608A.101P} demonstrated how taking a three-point finite difference approximation to the derivative of time series data alters the power spectrum. Namely, a $\sin^2\omega$ multiplying term is introduced to the power spectrum, where $\omega$ is an angular frequency varying between 0 and $\pi$ at the lowest and highest frequencies, respectively.
This means that the periodogram approaches zero towards these frequencies. Therefore, the first and final two frequency bins had to be removed from the analysis. Based on this, the power spectra shown in Figs.~\ref{fig:long spectra}--\ref{fig:short_spectra_2} were divided by the $\sin^2\omega$ term before the confidence limits were calculated.     

\subsection{Empirical mode decomposition (EMD) analysis}

Operating locally and not being restricted by the basis of expansion \citep[see][]{1998RSPSA.454..903E, 2008RvGeo..46.2006H}, EMD is found to be naturally suitable for processing QPP in solar and stellar flares \citep[see e.g.][]{2010PPCF...52l4009N, 2015A&A...574A..53K, 2016ApJ...830..110C, 2018MNRAS.tmp...77D}, including non-stationary QPP with a strong period drift \citep[cf. those studied by][]{2010SoPh..267..329K}. By the full analogy with the periodogram approach, where significance of the detected spectral components can be tested by the method discussed in Sec.~\ref{sec:fourier}, not all EMD modes necessarily correspond to statistically significant oscillatory processes. The latter must be checked with a similar significance test and taken as an intrinsic and compulsory feature of the method. The statistics of power-law distributed processes, including random noise as a specific case, was incorporated in the EMD analysis by \citet{2004RSPSA.460.1597W} and \citet{2016A&A...592A.153K}, and this is briefly outlined here. In this study, we treat the observational signal as a superposition of a smooth slowly varying trend (the flare itself), a possible oscillation that can be modulated by amplitude, period, and phase, and coloured noise characterised by the same power law index $\alpha$, as introduced in Sec.~\ref{sec:fourier}. In the parlance of the EMD analysis, the dependence of the total energy $E_\mathrm{m}$ of intrinsic modes upon the period $P_\mathrm{m}$ in pure coloured noise is given by
\begin{equation}
	\label{eq:col_noise_emd}
	E_\mathrm{m}P_\mathrm{m}^{1-\alpha}=\mathrm{const},
\end{equation}
which is similar to Eq.~(\ref{eq:col_noise_fourier}) describing behaviour of coloured noise in the Fourier power spectrum.

Having obtained the dependence of the total energy of identified intrinsic modes upon their mean period, which may be referred to as an EMD spectrum (see Figs.~\ref{fig:long spectra}--\ref{fig:short_spectra_2}), one can estimate the value of the power law index $\alpha$, i.e. the colour of noise superimposed on the initial observational signal, by approximation a functional form of Eq.~(\ref{eq:col_noise_emd}) into this spectrum. The very last mode of the expansion, showing an overall trend of the signal, is usually excluded from this approximation, corresponding to an effective and self-consistent detrending of the original light curve. Furthermore, given the definition of the total energy of an intrinsic mode as a sum of squares of its instantaneous amplitudes, it is therefore chi-squared distributed at each instantaneous period in the EMD spectrum. The parameter of the distribution function, the number of degrees of freedom, varies with the period and the colour of noise considered \citep[see][for details]{2016A&A...592A.153K}. This distribution can be visualised via the confidence interval of e.g. 99\% significance in the EMD spectrum. Thus the modes whose energies lie off this interval should be treated as statistically significant, while the modes within this interval are indistinguishable from the background processes governed by the power law with a certain value of the index $\alpha$, i.e. coloured noise.

In this study, we applied the described methodology to both the full flare light curve, as shown in Figs.~\ref{fig:flare} and \ref{fig:long spectra}, and to those trimmed sections of the flare which show the highest signal-to-noise ratio of spectral peaks in the periodogram-based analysis (see Figs.~\ref{fig:short_spectra_1} and \ref{fig:short_spectra_2}), focusing on the GOES data only.

\section{Results}
\label{sec:res}

\begin{deluxetable*}{cccccc}[t]
	\tablecaption{Statistically significant periods detected above 95\% significance level (except where indicated) in the signals shown in Figs.~\ref{fig:flare}, \ref{fig:short_spectra_1}, and \ref{fig:short_spectra_2}, with the periodogram and EMD methods. {The periodogram and EMD results are separated by a slash, \lq\lq /\rq\rq. When only one value appears, it was determined from the periodogram.}
		\label{table}
	}
	\tablecolumns{6}
	\tablenum{1}
	\tablewidth{0pt}
	\tablehead{
		\colhead{Start time (UTC)} &
		\colhead{End time (UTC)} &
		\colhead{QPP duration (s)} &
		\colhead{1--8\,\AA\ Period (s)} & \colhead{0.5--4\,\AA\ Period (s)} & \colhead{1--70\,\AA\ Period (s)} \\
		\colhead{} & \colhead{} & \colhead{}&
		\colhead{Periodogram/EMD} & \colhead{Periodogram/EMD} & \colhead{Periodogram}
	}
	\startdata
	11:55:01 & 12:14:58 &1197& $265^{+83}_{-63}$ & $254^{+71}_{-55}$(94\%)/ $276^{+43}_{-81}$& $293^{+102}_{-76}$ \\
	11:57:04 & 11:59:31 &147& $12.1^{+1.4}_{-1.3}$/$11.6^{+2.7}_{-2.0}$& $-$ & $11.7^{+2.3}_{-1.9}$ \\
	11:59:31 & 12:06:01 &390& $18.9^{+0.2}_{-0.2}$ & $18.3^{+4.8}_{-3.8}$ /$18.9^{+5.3}_{-3.9}$& $18.9^{+0.2}_{-0.2}$ \\
	12:06:03 & 12:08:04 &121& $21.2^{+0.8}_{-0.8}$ & $21.2^{+0.8}_{-0.8}$/$20.0^{+3.5}_{-2.6}$ & $20.3^{+0.4}_{-0.4}$ \\
	12:08:01 & 12:17:48 &587&$-$& $24.6^{+4.0}_{-3.4}$/$25.9^{+5.3}_{-6.3}$, $58.9^{+11.5}_{-9.4}$ & $-$ \\
	\enddata
\end{deluxetable*}

The oscillation periods found to be statistically significant by the periodogram and EMD analyses in the full and trimmed flare light curves are summarised in Table~\ref{table}. They can be attributed to two different types of QPP observed simultaneously in this flare: those with a non-stationary short period drifting from about 12 to 25 seconds, and the other with a much longer period varying from about 4 to 5 minutes in different observational wavebands. In addition, EMD detected a 1-min oscillation pronounced from 12:08:01 to 12:17:48\,UT (a post-flare phase), which is not found to be significant using the periodogram-based technique.

The evidence of the longer-period variability is shown in Fig.~\ref{fig:long spectra} by both the periodograms and EMD spectrum, with confidence above 95\%. It is detected in both the GOES/XRS and SDO/EVE/ESP observations, and at all analysed wavebands. Its behaviour in the time domain is shown in Fig.~\ref{fig:flare}. It represents a rapidly decaying oscillation with a harmonic shape and a relatively stable period of about 4 to 5 minutes. The top right panel of Fig.~\ref{fig:flare} appears to show that the oscillation is modulated by a Gaussian envelope, which was found to be substantially better than an exponential form, with a damping time of about 7.5 minutes. This value gives an oscillation quality factor (defined as the ratio of the damping time to the period) of about 1.6.

The spectra of the shorter flare sections containing statistically significant oscillatory components and their corresponding time series are shown in Figs.~\ref{fig:short_spectra_1} and \ref{fig:short_spectra_2}. Their oscillation periods are found to be similar by both the periodogram and EMD analyses, and gradually increase with the progression of the flare from about 12 seconds at the flare maximum to about 25 seconds in the flare decay phase (see Table~\ref{table}). Unlike the longer-period variation described above, the detected shorter-period oscillations have a rather intermittent, wave-train-like, amplitude modulation, with the oscillation power highly localised in time (see e.g. the oscillation profile found in the time interval from 12:06:03 to 12:08:04\,UT). In addition, EMD detected a 1-min oscillation in the decay phase section starting from 12:08:01\,UT, with the confidence above the 99\% level. It is shown in combination with the 25 seconds component and a low-frequency trend in Fig.~\ref{fig:short_spectra_2}, and has a similar amplitude modulated behaviour.

\section{Discussion and conclusions}
\label{sec:disc}

Our study demonstrates the presence of at least two QPPs in an X9.3 flare.  In contrast with the QPP detection in an X14.4 flare by \citet{2006A&A...460..865M}, the detected QPP occur in the thermal emission.
The observed values of the longer oscillation period (4--5 minutes) and quality factor (1.6) are consistent with those typically detected in so-called SUMER oscillations \citep[e.g.][]{2011SSRv..158..397W} that are usually interpreted as slow magnetoacoustic oscillations in flaring loops. Similar values were found in soft X-ray intensity QPP in both stellar and less powerful solar flares \citep{2016ApJ...830..110C}. In addition, the Gaussian damping is similar to the decaying oscillatory patterns in stellar flares observed in WL \citep{2016MNRAS.459.3659P}. {The observed period, quality factor and the fact that the thermal emission is modulated suggest that a possible mechanism for this QPP is a standing slow magnetoacoustic oscillation in the site of the analysed flare {(see the results of the numerical modelling by e.g. \citeauthor{2005A&A...436..701S} \citeyear{2005A&A...436..701S} and \citeauthor{2007ApJ...659L.173T} \citeyear{2007ApJ...659L.173T}). However, we cannot rule out the possibility that the QPP are associated with other mechanisms, such as repetitive reconnection.}

The other period, 12--25~s, is in the range of the most common QPP detected in solar flares \citep[see][]{2016ApJ...833..284I, 2018SSRv..214...45M}. QPP with periods in this range are usually interpreted as being associated with standing sausage oscillations of coronal loops. The period of this oscillation mode is determined by the parallel or perpendicular Alfv\'en transit time in the oscillating plasma structure in the trapped \citep[e.g.][]{1984ApJ...279..857R, 1982SvA....26..340Z} and leaky regimes \citep[e.g.][]{1986SoPh..103..277C, 2007AstL...33..706K, 2012ApJ...761..134N}, respectively. In this interpretation, the observed gradual increase of the oscillation period could be readily attributed to the gradual evolution of the physical parameters in the oscillating loop, {for example the increase in loop length \citep{2016ApJ...827L..30H}}, and/or the increase in plasma density because of the evaporation upflows. The detection of this kind of QPP in {such an energetic solar} flare suggests that similar QPP could be found in stellar flares too. 


The simultaneous occurrence of QPP with periods of several minutes and a few tens of seconds in an X9.3 solar flare further strengthens the conclusion that QPP are a common feature of powerful energy releases. Although solar QPP periods can range from sub-second to over a minute, the shorter period observed here is similar to the periods detected in many solar flares \citep[e.g. recent surveys by][]{2016ApJ...833..284I, 2017A&A...608A.101P}. The longer-period QPP has a period similar to {the majority of} those detected in stellar flares \citep[e.g.][]{2016ApJ...830..110C, 2016MNRAS.459.3659P}. This apparent difference between solar and stellar QPP periods is {probably artificial}, likely being due to a selection bias as on stars we only tend to observe more energetic, longer-lasting flares and, importantly, the observational time resolution is often coarser. Furthermore, many typical solar flares have shorter lifetimes, prohibiting the detection of longer-period QPPs. Here the detection of the longer-period QPP is only possible because the flare is unusually energetic and long-lived. The simultaneous detection of the two distinct QPP timescales in a powerful solar flare, whose energy is comparable to that of stellar flares, indicates that the longer- and shorter-period QPP regimes are not mutually exclusive, which could therefore indicate the similarity of physical mechanisms responsible for the energy releases on the Sun and stars.

\acknowledgments
AMB, DYK and VMN acknowledge the support of the STFC consolidated grant ST/L000733/1. This work was supported in part by the Russian Foundation for Basic Research grant No. 17-52-80064 (VMN). We also acknowledge support from the International Space Science Institute for the team \lq\lq Quasi-periodic Pulsations in Stellar Flares: a Tool for Studying the Solar-Stellar Connection\rq\rq. VMN acknowledges the Russian Foundation for Basic Research grant No. 17-52-80064.

\bibliographystyle{aasjournal} 



\end{document}